\documentclass[conference, 10pt, A4]{IEEEtran}
\IEEEoverridecommandlockouts

\usepackage[font=footnotesize]{caption}
\usepackage{setspace}
\usepackage{enumitem}
\usepackage{amsmath} 
\usepackage{cite}
\usepackage{amsmath,amssymb,amsfonts}
\usepackage{algorithm}
\usepackage{algpseudocode}
\usepackage{algorithmicx}
\usepackage{graphicx}
\usepackage{textcomp}
\usepackage{caption}
\usepackage{subcaption}
\usepackage{xcolor}

\usepackage{enumitem}
\usepackage[font=footnotesize]{caption}
\usepackage{tikz}

\usepackage{fancyhdr}
\newcommand*\circled[1]{\tikz[baseline=(char.base)]{
            \node[shape=circle,fill,inner sep=1pt] (char) {\textcolor{white}{#1}};}}

\renewcommand{\IEEEbibitemsep}{0.5pt}
\makeatletter
\IEEEtriggercmd{\reset@font\normalfont}
\makeatother
\IEEEtriggeratref{1}
    
\begin{document}

\title{FAQ: Mitigating the Impact of Faults in the Weight Memory of DNN Accelerators through \\ Fault-Aware Quantization}

\author{\IEEEauthorblockN{Muhammad Abdullah Hanif, Muhammad Shafique}
\IEEEauthorblockA{\textit{Division of Engineering, New York University
Abu Dhabi (NYUAD), Abu Dhabi, United Arab Emirates}\\
mh6117@nyu.edu, muhammad.shafique@nyu.edu}}

\maketitle
\thispagestyle{fancy}
\begin{abstract}

Permanent faults induced due to imperfections in the manufacturing process of Deep Neural Network (DNN) accelerators are a major concern, as they negatively impact the manufacturing yield of the chip fabrication process. Fault-aware training is the state-of-the-art approach for mitigating such faults. However, it incurs huge retraining overheads, specifically when used for large DNNs trained on complex datasets. To address this issue, we propose a novel Fault-Aware Quantization (FAQ) technique for mitigating the effects of stuck-at permanent faults in the on-chip weight memory of DNN accelerators at a negligible overhead cost compared to fault-aware retraining while offering comparable accuracy results. We propose a lookup table-based algorithm to achieve ultra-low model conversion time. We present extensive evaluation of the proposed approach using five different DNNs, i.e., ResNet-18, VGG11, VGG16, AlexNet and MobileNetV2, and three different datasets, i.e., CIFAR-10, CIFAR-100 and ImageNet. The results demonstrate that FAQ helps in maintaining the baseline accuracy of the DNNs at low and moderate fault rates without involving costly fault-aware training. For example, for ResNet-18 trained on the CIFAR-10 dataset, at 0.04 fault rate FAQ offers (on average) an increase of 76.38\% in accuracy. Similarly, for VGG11 trained on the CIFAR-10 dataset, at 0.04 fault rate FAQ offers (on average) an increase of 70.47\% in accuracy. The results also show that FAQ incurs negligible overheads, i.e., less than 5\% of the time required to run 1 epoch of retraining. We additionally demonstrate the efficacy of our technique when used in conjunction with fault-aware retraining and show that the use of FAQ inside fault-aware retraining enables fast accuracy recovery. For example, for ResNet18 trained on the CIFAR-100 dataset, at 0.1 fault rate retraining without FAQ enables (on average) 4.36\% accuracy recovery while retraining with FAQ enables (on average) 66.97\% accuracy recovery compared to no mitigation approach. 

\end{abstract}

\section{Introduction}

Deep Neural Networks (DNNs) are state of the art for a wide spectrum of complex AI problems nowadays~\cite{lecun2015deep}~\cite{sarker2021deep}. 
However, the computational requirements of these high-accuracy DNNs restrict their deployment in resource-constrained scenarios~\cite{shafique2018overview, shafique2021tinyml, sze2017efficient}. 
To enable efficient DNN inference, specifically at the edge, specialized accelerators such as TPU~\cite{jouppi2017datacenter}, MAERI~\cite{kwon2018maeri} and Eyeriss~\cite{7738524, 7551407, chen2019eyeriss} are used~\cite{capra2020hardware, capra2020updated, marchisio2019deep}. 
These accelerators are manufactured using advanced nano-scale CMOS technology to achieve high efficiency gains. A key challenge associated with nano-scale CMOS devices is that they face various reliability threats such as soft errors, aging, process variations and permanent faults~\cite{hanif2018robust}. 
These faults usually manifest as bit flips in the system, and earlier works have shown that even a single bit-flip error at a critical location in the system can significantly degrade the accuracy of DNNs~\cite{pattabiraman2020error}. 

\begin{figure}[t]
\centering
\includegraphics[width=1\linewidth]{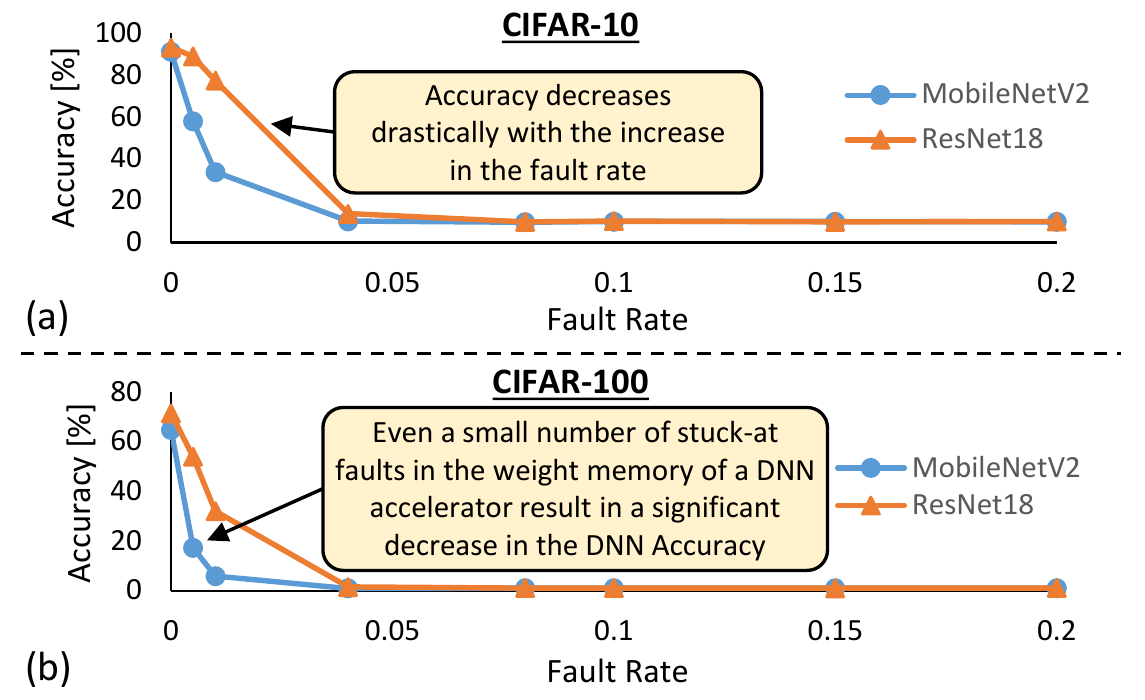}
\caption{Impact of stuck-at faults in the on-chip weight memory of a DNN accelerator on the accuracy of different DNNs. (a) Results for DNNs trained on the Cifar-10 dataset. (b) Results for DNNs trained on the Cifar-100 dataset.}
\label{fig:Intro_results}
\end{figure}

Permanent faults induced due to imperfections in the chip fabrication process (manufacturing defects) are a major concern, as they impact the manufacturing yield and, thereby, the cost of devices. 
These faults manifest as stuck-at bits at the hardware level, i.e., lines that always carry the logical signal “0” or “1” as the result of a short or open circuit~\cite{WADDEN2017617}. 
Our experimental analysis in Fig.~\ref{fig:Intro_results} highlights the impact of stuck-at faults in the on-chip weight memory of a DNN accelerator on the accuracy of different DNNs trained for image classification application. 
Fig.~\ref{fig:Intro_results} illustrates that the accuracy of DNNs decreases drastically with the increase in the fault rate, and therefore, hardware having permanent faults cannot be used for reliable DNN inference. 

\begin{figure}[tb]
\centering
\includegraphics[width=1\linewidth]{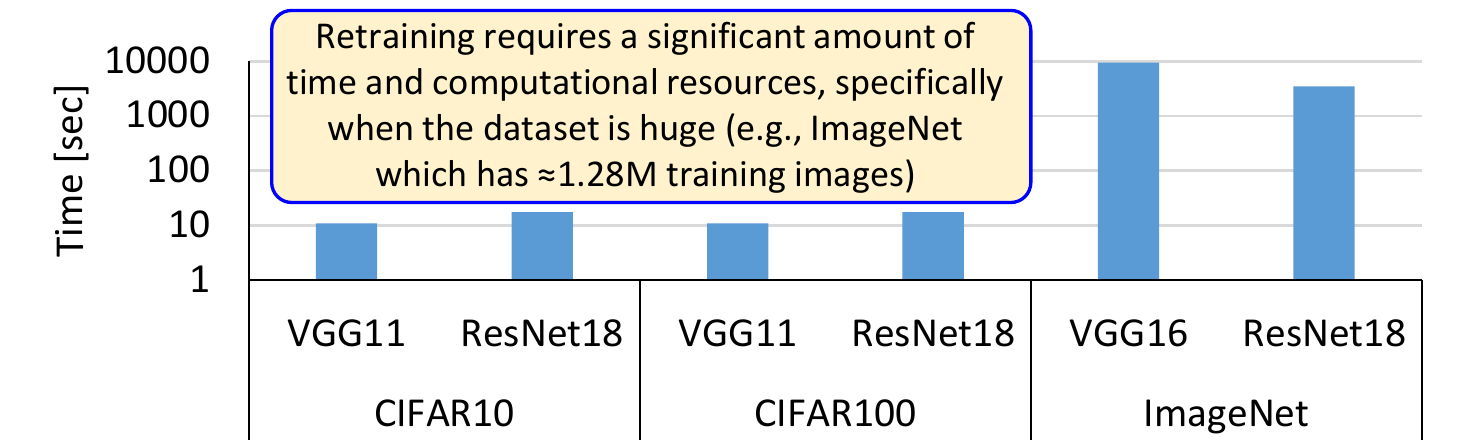}
\caption{Time for one epoch of retraining for different DNN and dataset pairs.}
\label{fig:Motivational_Figure_1}
\end{figure}

\textbf{State of the Art and their Limitations:} A number of works in the literature have focused on addressing permanent faults in DNN accelerators. 
For example, Zhang et al.~\cite{zhang2018analyzing}\cite{zhang2019fault} proposed Fault-Aware Pruning (FAP) and FAP~+~Training (FAP+T) for addressing permanent faults in the computational array of a DNN accelerator. 
However, such works only target faults in the computational array of DNN accelerators. 
Permanent faults can also occur in the on-chip memory of DNN accelerators. 
Various techniques have been proposed to address faults in the memory of DNN accelerators. 
These techniques can be divided into two major categories: (1) Retraining-based techniques and (2) Correction-based techniques. 
The retraining-based techniques mainly employ fault-aware training to mitigate the effects of faults~\cite{zahid2020fat, koppula2019eden, kim2018matic}. However, such techniques lead to \textit{huge retraining overheads} (see Fig.~\ref{fig:Motivational_Figure_1}), rendering them infeasible for many practical settings. Moreover, some of these techniques are not particularly designed to address permanent faults, e.g.,~\cite{koppula2019eden}. 
On the other hand, correction-based techniques, such as the one proposed in~\cite{reagen2016minerva} and Error Correction Codes (ECC)~\cite{levine1976special}, employ additional hardware components for error detection and mitigation, which (1) require hardware architecture modifications and (2) lead to run-time overheads. 

\textbf{Targeted Research Problem:} To overcome the limitations of the state-of-the-art works, in this paper, we focus on studying how the intrinsic resilience of DNNs to errors can be exploited to build a low-cost technique for mitigating the effects of stuck-at faults in the on-chip weight memory of DNN accelerators. 

\begin{figure}[h]
\centering
\includegraphics[width=1\linewidth]{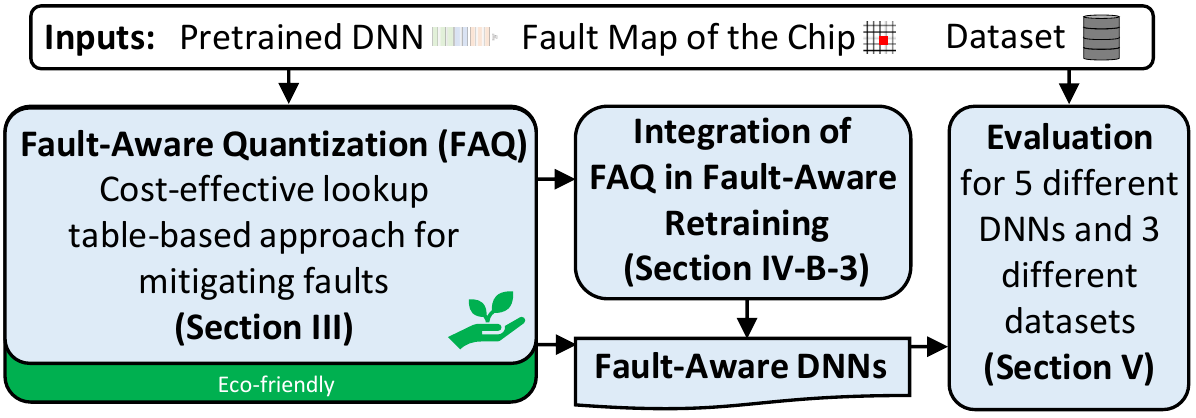}
\caption{Overview of our novel contributions.\vspace{-10pt}}
\label{fig:contributions}
\end{figure}

\textbf{Our Novel Contributions:} 
To address the above problem, we propose a novel Fault-Aware Quantization (FAQ) technique for mitigating the effects of stuck-at permanent faults in the on-chip weight memory of DNN accelerators. Towards this, the key novel contributions of this paper (highlighted in Fig.~\ref{fig:contributions}) are: 

\begin{enumerate}
    \item \textbf{Fault-Aware Quantization (FAQ):} A low-cost technique for mitigating the effects of stuck-at permanent faults in the on-chip weight memory of a DNN accelerator. 
    The technique exploits the static nature of stuck-at permanent faults and the capability of the weight memory to reproduce some of the values correctly even in the presence of stuck-at faults to offer low-cost fault mitigation.  
    \item \textbf{Efficient Implementation:} We introduce a lookup table-based algorithm for time and resource efficient execution of the proposed FAQ technique. 
    \item \textbf{Effectiveness for Fault-Aware Retraining:} We demonstrate the effectiveness of our proposed FAQ technique for accelerating accuracy recovery (towards the baseline) during fault-aware retraining. 
    \item \textbf{Extensive Evaluation:} We present extensive result using five different DNNs, i.e., ResNet-18, VGG11, VGG16, AlexNet and MobileNetV2, and three different datasets, i.e., CIFAR-10, CIFAR-100 and ImageNet.  
\end{enumerate}

Our results highlight that the proposed FAQ technique helps in maintaining the baseline accuracy of the DNNs at low and moderate fault rates without involving costly fault-aware (re)training. For example, at 0.04 fault rate, for ResNet18 trained on the CIFAR-10 dataset, FAQ offers (on average) an increase of 76.38\% in accuracy, and for VGG11 trained on the CIFAR-10 dataset, FAQ offers (on average) an increase of 70.47\% in accuracy. Also, the model conversion incurs negligible overheads, i.e., less than 5\% of the time required to run 1 epoch of retraining. The results also demonstrate that FAQ can be employed to improve the effectiveness of fault-aware retraining approaches. For example, for ResNet18 trained on the CIFAR-100 dataset, at 0.1 fault rate, retraining without FAQ enables (on average) 4.36\% accuracy recovery while retraining with FAQ enables (on average) 66.97\% accuracy recovery compared to no mitigation approach. 

\section{Preliminaries}

\subsection{Deep Neural Networks (DNNs)}
A DNN is an interconnected network of neurons, where each neuron performs a weighted sum operation, and then applies a non-linear activation function to the output. The functionality of a neuron can be written as $O = F(\sum_i W_i * A_i + b)$, where $O$ represents the output, $W_i$ represents the $i^{th}$ weight, $A_i$ represents the $i^{th}$ input activation, $b$ represents the bias, and $F$ represents the non-linear activation function. These neurons are usually arranged in the form of layers, and the layers are then connected to form a DNN. 

\subsection{Convolutional Neural Networks (CNNs)}

\begin{figure}[ht]
\centering
\includegraphics[width=1\linewidth]{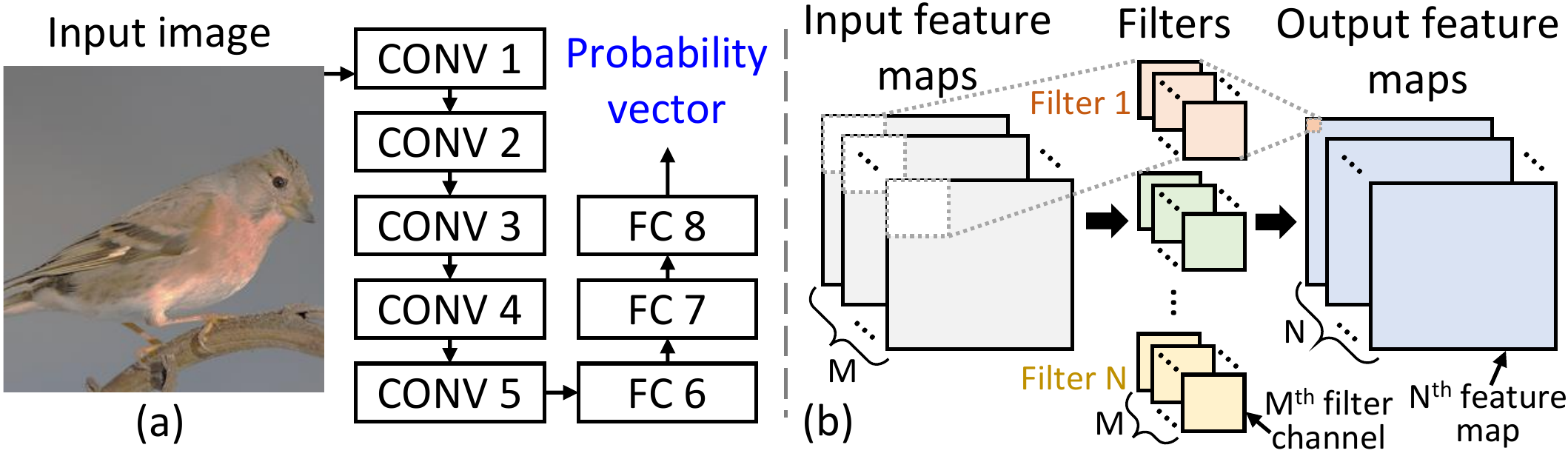}
\caption{(a) Architecture of the AlexNet network composed of five convolutional layers and three fully-connected layers designed for image classification application. (b) Detailed illustration of a convolutional layer showing convolution of input feature maps with filters to generate output feature maps.\vspace{-10pt}} 
\label{fig:2}
\end{figure}

CNNs are a special type of DNNs designed to process spatially correlated data such as images~\cite{lecun2015deep}. They are mainly composed of convolutional (CONV) layers and fully-connected (FC) layers, as shown in Fig.~\ref{fig:2}a. 
CONV layers are used to extract features from the input while FC layers are used at the end of the network for classification. An overview of a CONV layer composed of $N$ number of filters is shown in Fig.~\ref{fig:2}b. 

\subsection{DNN Accelerators}
\label{sec:DNN_Accelerators}

\begin{figure}[ht]
\centering
\includegraphics[width=1\linewidth]{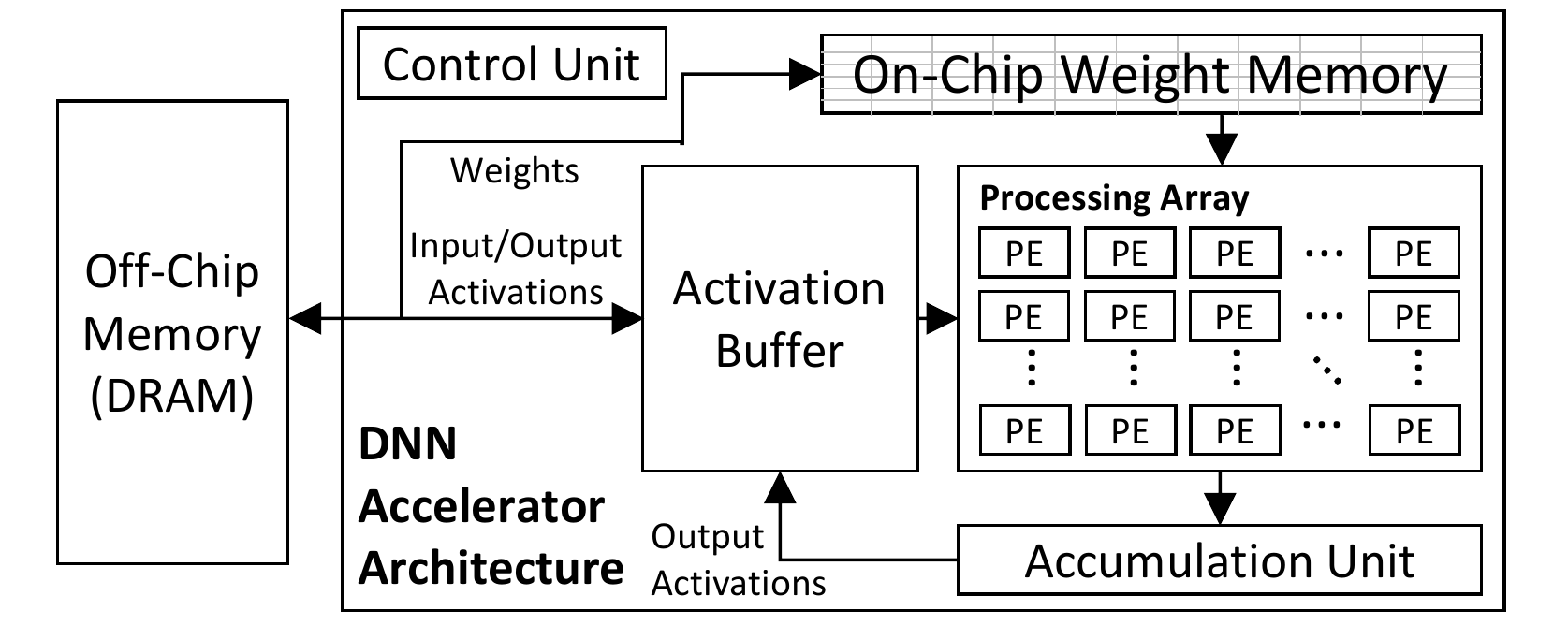}
\caption{Overview of a systolic array-based DNN hardware accelerator illustrating the main components involved in the design.\vspace{-10pt}}
\label{fig:HA1}
\end{figure}

\begin{figure}[ht]
\centering
\includegraphics[width=1\linewidth]{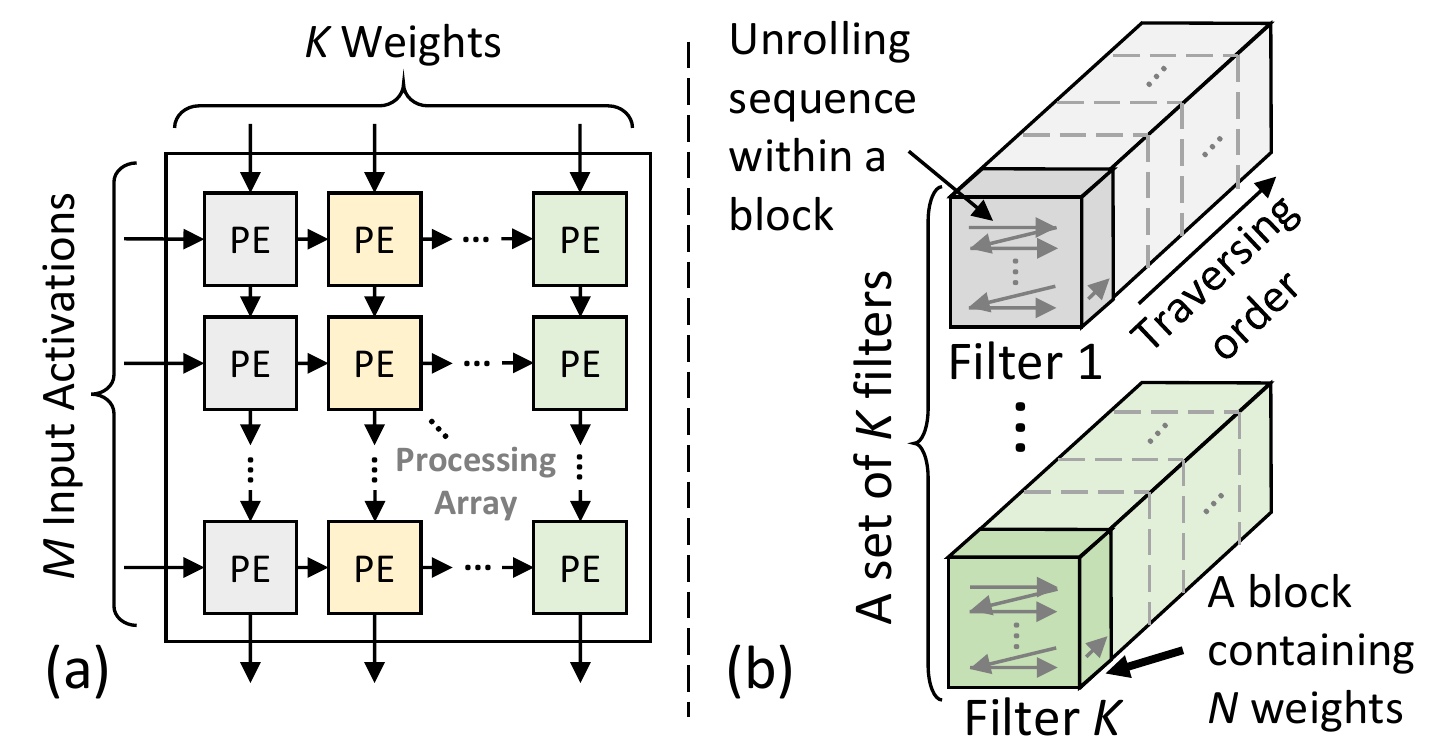}
\caption{(a) Detailed view of a processing array used to perform matrix multiplication operations. This design follows a weight stationary dataflow where weights are mapped onto the processing array and are kept stationary while corresponding activations are passed through horizontal channels to perform MAC operations. (b) Dataflow illustrating the sequence in which filter weights are accessed and mapped onto the processing array shown in (a) to perform corresponding MAC operations.\vspace{-5pt}}
\label{fig:HA2}
\end{figure}

DNN accelerators are commonly used for efficient execution of DNN workloads~\cite{jouppi2017datacenter}\cite{kwon2018maeri}\cite{chen2019eyeriss}.
As illustrated in Fig.~\ref{fig:HA1}, a DNN accelerator is mainly composed of a processing array and on-chip memory buffers. 
Fig.~\ref{fig:HA2}a shows a detailed view of a processing array similar to the one used in~\cite{jouppi2017datacenter}, \cite{zhang2018analyzing}, \cite{zhang2019fault}, and~\cite{abdullah2020salvagednn}. 
For performing dot-product operations, the array is first loaded with weights through vertical channels. 
The weights are then kept stationary inside the array while the activations (inputs) are fed through the horizontal channels. 
The generated partial sums flow downstream. 
Fig.~\ref{fig:HA2}b shows the dataflow for processing CONV as well as FC layers using the array in Fig.~\ref{fig:HA2}a. 
As shown in Fig.~\ref{fig:HA2}b, larger filters are divided into smaller segments based on the size of the array.
These segments are then moved to the on-chip weight memory and mapped to the array one by one to perform the corresponding operations. 
Note that only the weights belonging to the same filter/neuron can be mapped to the same column in the array shown in Fig.~\ref{fig:HA2}a.  

\section{Fault-Aware Quantization}
\label{sec:analysis_and_strategies}

\subsection{Concept Overview}

\begin{figure}[h]
\centering
\includegraphics[width=1\linewidth]{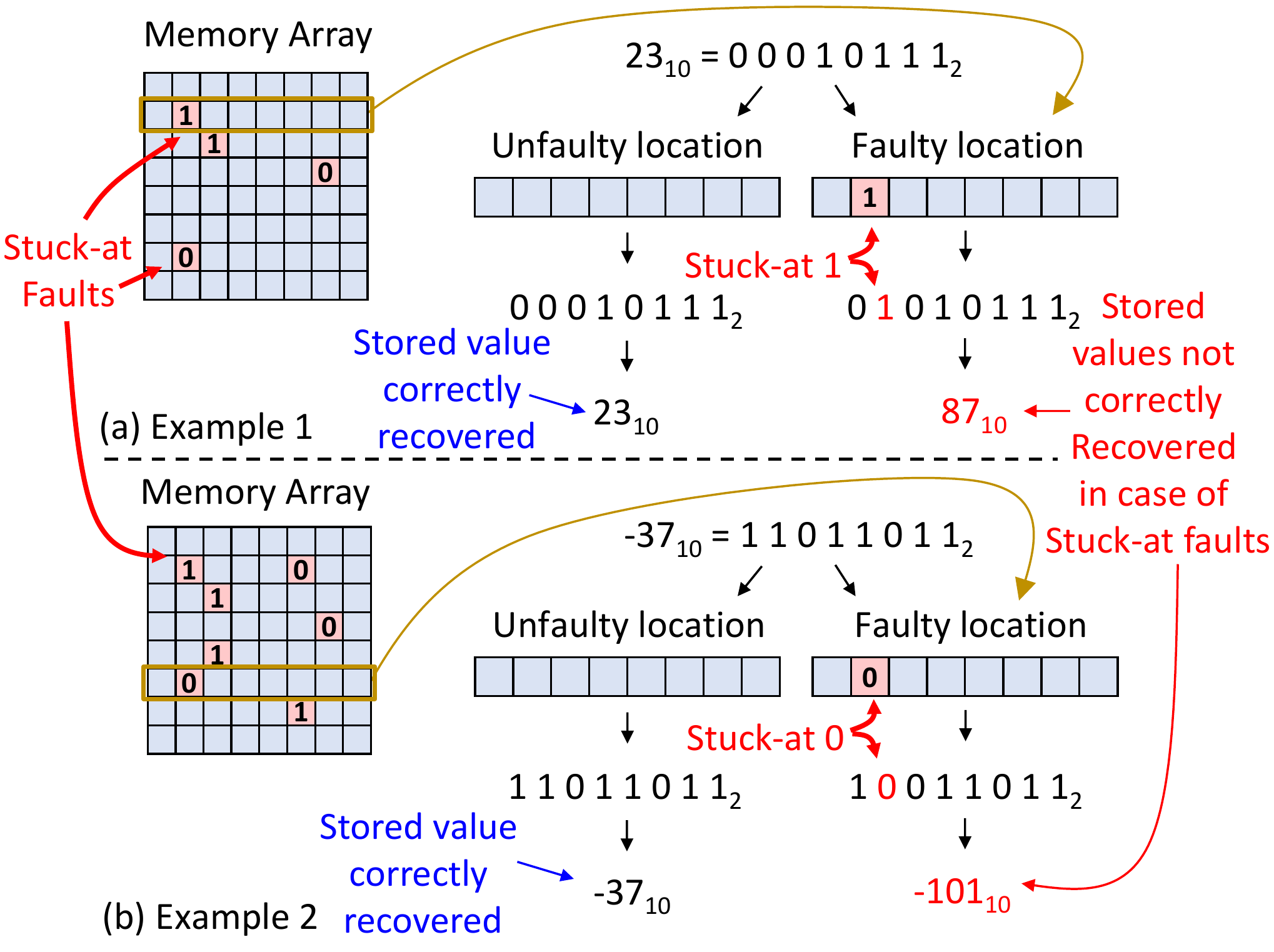}
\caption{Examples of the impact of stuck-at faults in the weight buffer on the stored values. (a) Impact of a stack-at 1 at the $7^{th}$ bit location on an 8-bit signed value. (a) Impact of a stack-at 0 on an 8-bit signed value.}
\label{fig:Problem_Example}
\end{figure}

Stuck-at faults in the on-chip weight memory of a DNN accelerator can significantly change the stored weight values and thereby have the potential to drastically reduce the accuracy of the executing DNN. 
Fig.~\ref{fig:Problem_Example} illustrates how a stuck-at fault at the $7^{th}$ bit location can change an 8-bit signed integer from $23_{10}$ to $87_{10}$ and $-37_{10}$ to $-101_{10}$. 
As on-chip weight memory is shared for weights, a single stuck-at fault in one memory cell can reflect in multiple weight values. 
Moreover, depending on the significance of the faulty location and the sensitivity of the affected weight, even a single bit-flip can drastically reduce the accuracy of the executing DNN~\cite{pattabiraman2020error}. 

\begin{figure}[ht]
\centering
\includegraphics[width=1\linewidth]{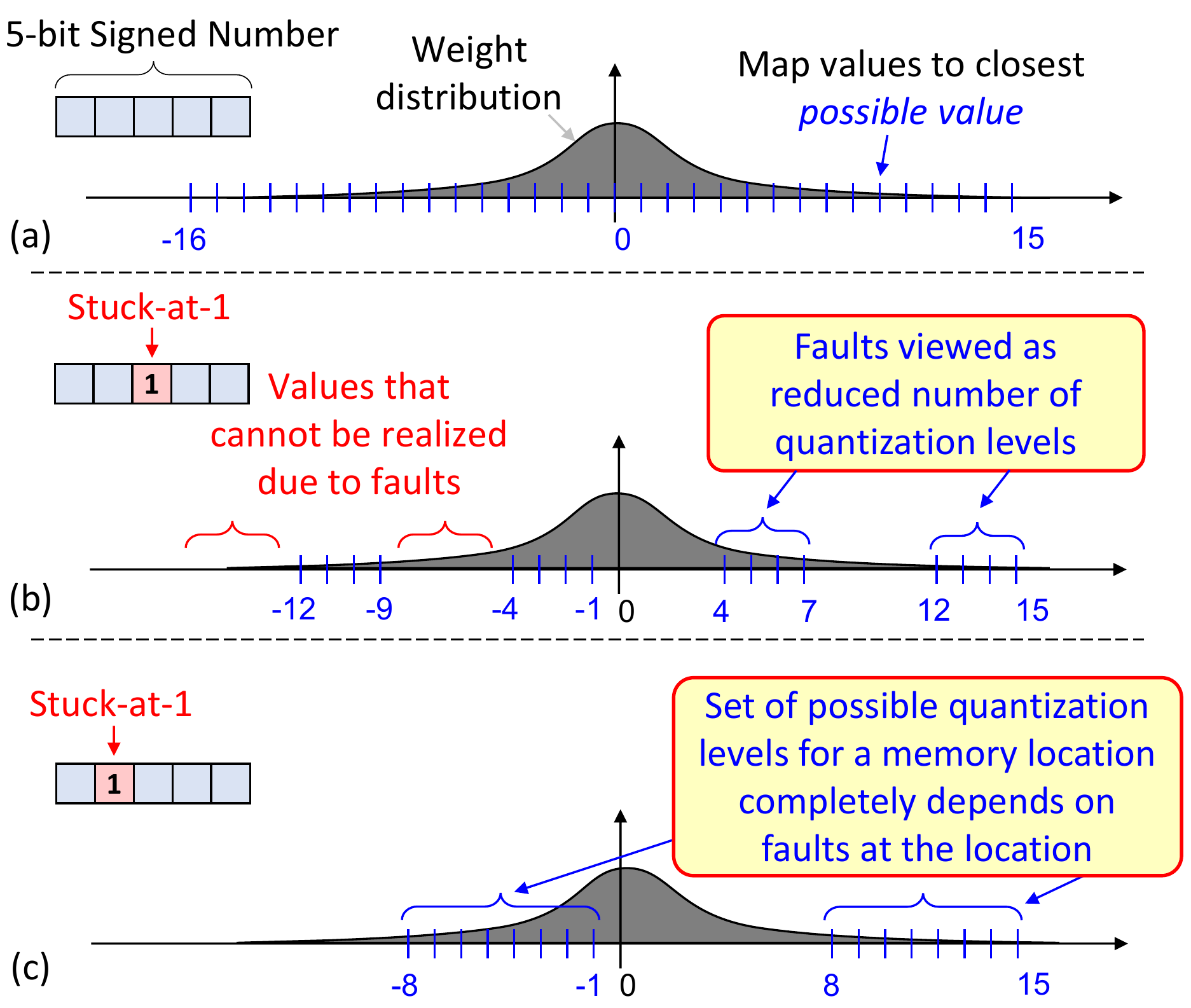}
\caption{Fault-Aware Quantization (FAQ): A faulty memory location can be viewed as a location that can reproduce only a limited set of values correctly. (a) Quantization levels for a 5-bit signed number format. (b) Stuck-at 1 at $3^{rd}$ bit location reduces the set of values that can be realized correctly. (c) Stuck-at 1 at $4^{th}$ bit location lead to a different set of correctly realizable values than the ones in (b).}
\label{fig:Concept}
\end{figure}

In this work, we highlight that a faulty memory location that has stuck-at-1 and/or stuck-at-0 faults can be viewed as a location that can reproduce only a limited set of values correctly. 
This concept is illustrated with the help of Fig.~\ref{fig:Concept} for 5-bit memory locations. 
The figure shows that depending on the location and type of faults, some values cannot be realized. 
Moreover, mapping values that cannot be reproduced correctly to corresponding faulty locations lead to errors, which can be significant as shown earlier in Fig.~\ref{fig:Problem_Example}.  
Therefore, based on the above concept, we observe that the error due to stuck-at faults in memory can be reduced by quantizing weights to the nearest value that can be correctly reproduced even in the presence of faults. 
Note that as each location can have a different fault pattern (independent of other locations), the set of values that are valid for one location may not be valid for another, as can be observed from Figs.~\ref{fig:Concept}b and~\ref{fig:Concept}c. 
Therefore, each memory location can have a different set of valid (correctly reproducible) values, which is completely based on the memory fault map. 
To address the above challenges, in this work, we propose to: (1) compute the set of valid values for each possible fault pattern for the given data representation format; (2) for each fault pattern, define closest valid value for each value in the range of the given data representation format; (3) use the provided fault map and data mapping policy to define the valid values for each weight/parameter in the DNN; and (4) use the information from the above steps to efficiently update the weights of the DNN. 

\subsection{Fault-Aware Quantization Methodology} 

\begin{figure}[t]
\centering
\includegraphics[width=1\linewidth]{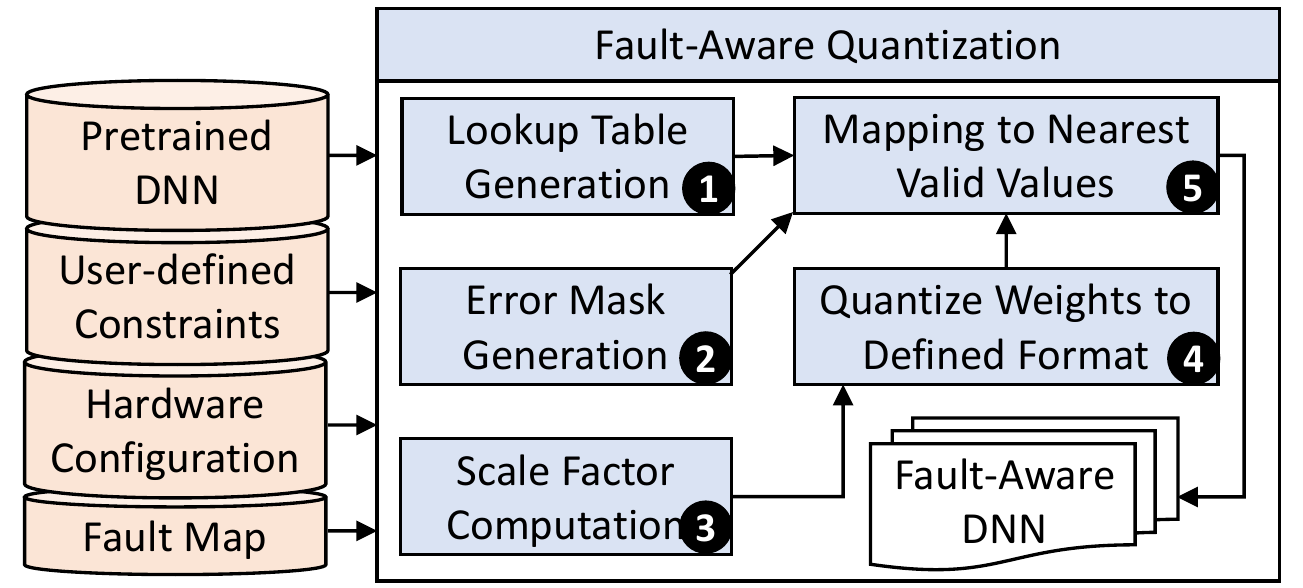}
\caption{Overview of our proposed FAQ methodology.}
\label{fig:methodology}
\end{figure}

Fig.~\ref{fig:methodology} presents an overview of our proposed methodology, which takes a pretrained DNN, user-defined constraints (i.e., bitwidth and data representation format for weights), hardware configuration (i.e., memory size and dataflow policy) and fault map as inputs, and generates a fault-aware DNN for the given fault map. 
The steps of the methodology are as follows: 

\begin{itemize}
\item In Step~\circled{1}, first, based on the user-defined bitwidth and data representation format, all the fault patterns are identified. Then, for each fault pattern, valid values that can correctly be reproduced are computed. Then, using these computed values, for each \{\textit{fault pattern}, \textit{possible value in the range of the given data representation format}\} pair, the closest possible valid value (that can correctly be reproduced) is computed. These values are then stored in a lookup table format for efficient retrieval. The resultant lookup table is then later used in Step~\circled{5} for adjusting the weights of the given DNN such that it results in minimal accuracy drop when subjected to stuck-at faults in the weight memory of the given DNN accelerator. The lookup table generation process is summarized in Algorithm~\ref{Algo:Lookup_Table}. 
    \item In Step~\circled{2}, using the given fault map and hardware configuration, fault pattern for each weight value is identified. The hardware configuration together with dataflow policy defines the memory location for each DNN weight (assuming a fixed mapping policy). The location together with the fault map defines the fault pattern a weight experiences during execution. This step results in an error mask which has information about the fault pattern for each individual weight in the given DNN. 
    \item In Step~\circled{3}, given the bitwidth and data representation format, scale factor is computed for each data-structure (weight and activation tensor) of the given DNN. These scale factors are then forwarded to Step~\circled{4}.
    \item In Step~\circled{4}, using the scale factors and the input information, the weights of the DNN are quantized to the user-defined data representation format without incorporating information of the faults in the system. The quantized values are then forwarded to Step~\circled{5}. 
    \item In Step~\circled{5}, the quantized values from Step~\circled{4} and the error mask from Step~\circled{2} are used to define query values. These values are then used to extract corresponding nearest valid values from the lookup table, which are then used as the final quantized values for the weights. 
\end{itemize}

\begin{algorithm}[ht]
\scriptsize	
\caption{Lookup Table Generation}
\label{Algo:Lookup_Table}
\begin{algorithmic}[1]
\Statex \textbf{Inputs:} bitwidth = 8; number\_format = Signed 2's complement format
\Statex \textbf{Outputs:} Lookup\_table \textcolor{gray}{\# Comment: Table that defines the nearest possible value for each fault pattern and value in the range of the given number representation format}
\Statex \textbf{Initialize:} Lookup\_table = Zeros(3$^{bitwidth}$, 2$^{bitwidth}$) 
\For i in [0, 1, 2, …, 3$^{bitwidth}$]: \textcolor{gray}{\# Comment: For each possible fault pattern}
\State Faults = Decimal2Ternary(i, bitwidth) \textcolor{gray}{\# Comment: Define fault pattern by converting i to ternary format having as many digits as the given bitwidth. Here, in the ternary format, `0' represents no fault, `1' represents stuck-at 0 fault and `2' represents stuck-at 1 fault.}
\State Range = [-2$^{bitwidth-1}$, -2$^{bitwidth-1}$+1, …, 2$^{bitwidth-1}$-1]
\State Range\_b = Convert2Binary(Range, number\_format, bitwidth)
\For j in [0, 1, 2, …, bitwidth]:
\If {Faults[j] $>$ 0} \textcolor{gray}{\# Comment: if j$^{th}$ location is faulty} 
\State Range\_b[:, j] = Range\_b[:, j] * 0 \textcolor{gray}{\# Comment: Mask the corresponding bits to 0 assuming the fault is stuck-at 0 fault}
\If {Faults[j] $==$ 2} \textcolor{gray}{\# Comment: Check if the fault is stuck-at 1 fault} 
\State Range\_b[:, j] = Range\_b[:, j] + 1 \textcolor{gray}{\# Comment: Mask the corresponding bits to 1 if the fault is stuck-at 1 fault}
\EndIf
\EndIf
\EndFor
\State Range\_dec\_Masked = Convert2Decimal(Range\_b, number\_format, bitwidth)
\State unique\_values = unique(Range\_dec\_Masked) \textcolor{gray}{\# Comment: find unique values}
\State unique\_value\_mat = unique\_values.repeat(repeats = [2$^{bitwidth}$], axis = 0) \textcolor{gray}{\# Comment: Generate unique values matrix by repeating unique\_values vector 2$^{bitwidth}$ times.}
\State relative\_distance = Range.reshape(2$^{bitwidth}$, 1).repeat(repeats = [unique\_value\_mat.shape[1]], axis = 1) - unique\_value\_mat \textcolor{gray}{\# Comment: Find distance of each value in the Range to each value in unique\_values}
\State indexes = Abs(relative\_distance).argmin(axis=1) \textcolor{gray}{\# Comment: Find indexes Store nearest possible values in the lookup table at corresponding locations}
\State Lookup\_table[i,:] = unique\_values[indexes] \textcolor{gray}{\# Comment: Store nearest possible values in the lookup table at corresponding locations}
\EndFor
\State Lookup\_table = Lookup\_table.reshape(-1) \textcolor{gray}{\# Comment: Convert to 1D vector}
\State \Return \textbf{Lookup\_table}
\end{algorithmic}
\end{algorithm}

\section{Results and Discussion}

\subsection{Experimental Setup}

\begin{figure}[ht]
\centering
\includegraphics[width=1\linewidth]{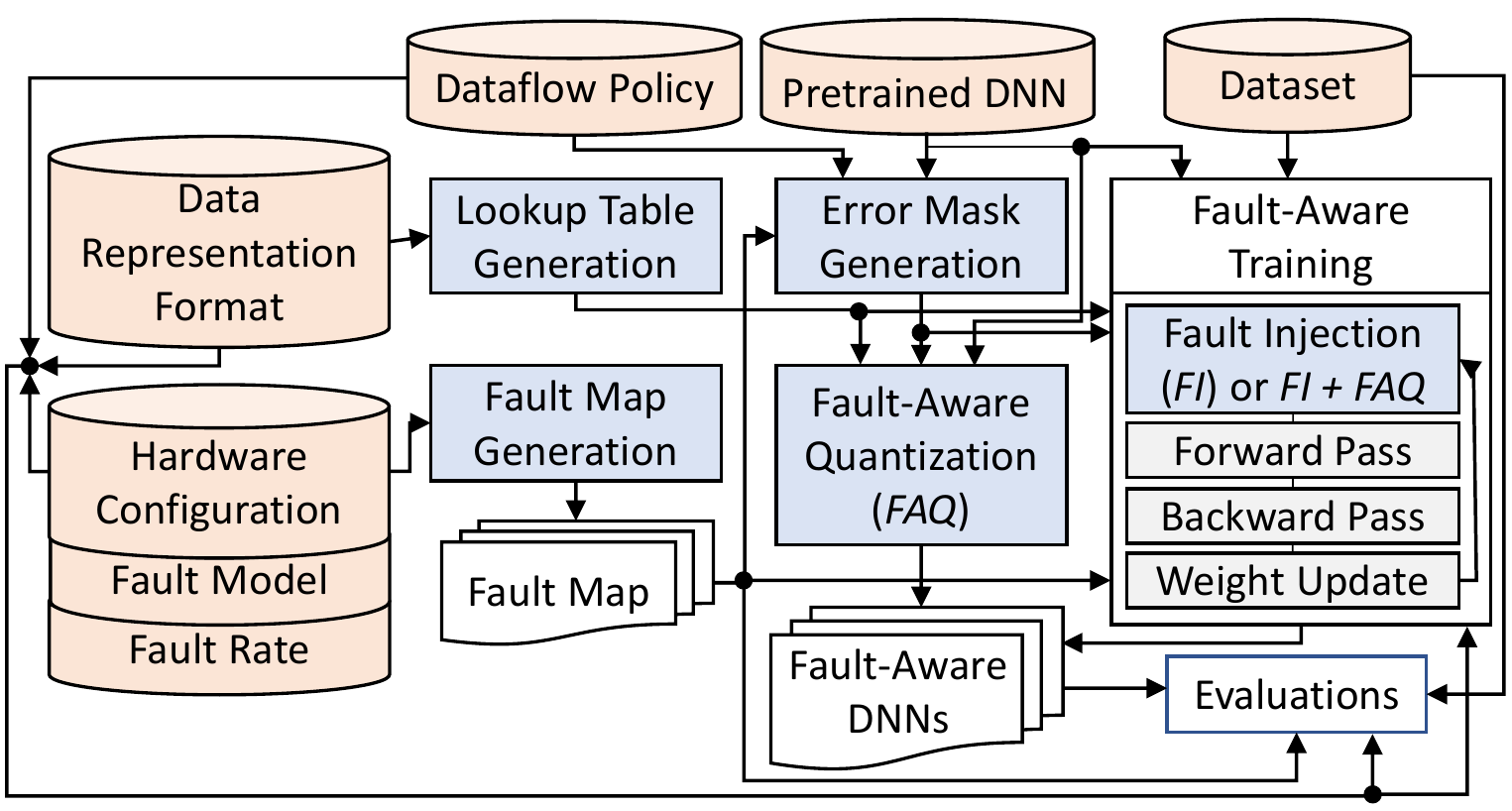}
\caption{An illustrative overview of the experimental setup}
\label{fig:Exp_Setup}
\end{figure}

To evaluate the proposed technique, we build an evaluation framework using the PyTorch library. The overview of the framework is illustrated in Fig.~\ref{fig:Exp_Setup}. As can be seen in the figure, the framework is composed of lookup table generation, fault-map generation, error mask generation, FAQ, fault-aware training and evaluation blocks. 
First, the lookup table generation block takes the desired data representation format (e.g., 8-bit signed fixed-point representation) and generates a table that defines the nearest possible value for each possible fault pattern and input value pair. 
Then, the fault-map generation block generates a fault map based on the provided hardware configuration (mainly on-chip weight memory size) and fault rate. For, the evaluations in this work, we assumed random stuck-at fault injection model, i.e., we use the provided fault rate as the probability of occurrence of a stuck-at fault in each individual on-chip weight memory cell, and then define the polarity of the stack-at fault based on an equiprobable event. 
The generated fault map (which corresponds to a faulty chip) is then forwarded to error mask generation block, which, considering the given dataflow policy, defines the faults each individual DNN weight/parameter will experience during execution on the given faulty chip. 
The error mask together with the lookup table is then used in FAQ block to modify the weight values. 
The generated fault-aware DNN is then used for final evaluation.

To show the effectiveness of the proposed methodology, we tested the proposed FAQ technique for five different DNNs, i.e., ResNet-18, VGG11, VGG16, AlexNet and MobileNetV2 VGG16 models. Moreover, we considered three different datasets, i.e., CIFAR-10, CIFAR-100 and ImageNet datasets. The considered dataset and DNN combinations are summarized in Table.~\ref{tab:Datasets_DNNs}. 
We performed evaluations for both with and without fault-aware retraining cases in order to highlight the significance of FAQ even for fault-aware retraining. 

\begin{table}[ht]
    \centering
    \caption{Datasets and DNNs considered for evaluation}
    \label{tab:Datasets_DNNs}
    \includegraphics[width=0.7\linewidth]{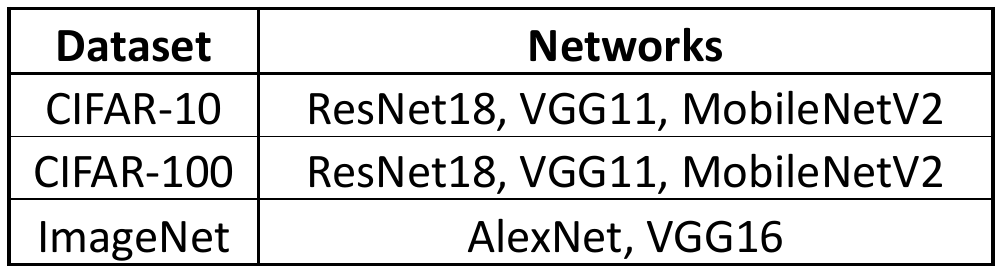}
\end{table}

For hardware configuration, we considered a TPU-like architecture presented in~\cite{zhang2018analyzing}. 
We assumed on-chip weight memory of size $256\times256$ bytes, where one block of weights is mapped at a time based on the dataflow explained in~\cite{zhang2018analyzing}. 
Moreover, for data representation format, we considered 8-bit signed fixed-point representation, as 8-bit quantization is wide used in embedded deep learning community. 

\subsection{Results}

\begin{figure*}[ht]
\centering
\includegraphics[width=1\linewidth]{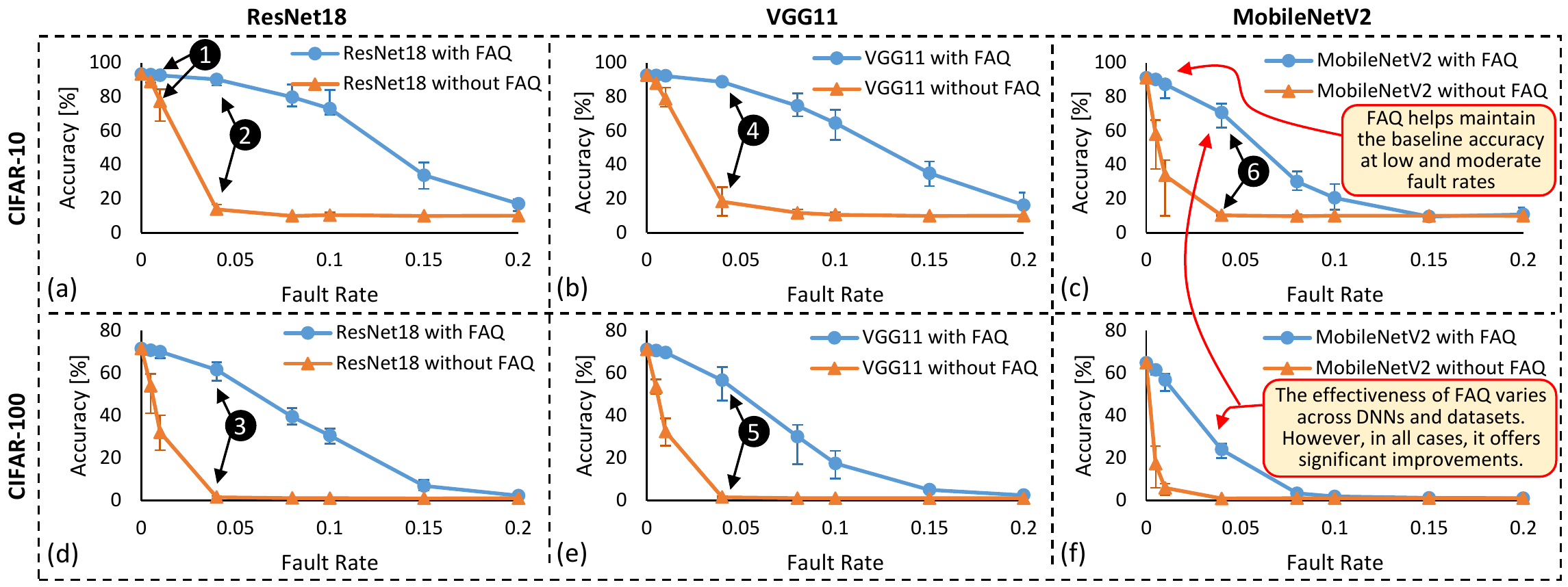}
\caption{Impact of FAQ on the accuracy of different DNNs when subjected to stuck-at faults in the on-chip weight memory of DNN accelerators. The error bars around each data point shows the minimum and maximum values observed in five independent experiments performed using different seed values.}
\label{fig:Results_1}
\end{figure*}

\subsubsection{Impact of FAQ on DNN Accuracy} 
Fig.~\ref{fig:Results_1} presents the impact of FAQ on the accuracy of DNNs subjected to stuck-at faults in the on-chip weight memory of DNN accelerators. 
The figure shows that for three different DNNs, i.e., ResNet18, VGG11 and MibileNetV2, and two different datasets, i.e., CIFAR-10 and CIFAR-100, the proposed FAQ technique helps in significantly increasing the accuracy of DNNs when executed using faulty hardware. 
Specifically, at low and moderate fault rates, it helps in maintaining close to the baseline accuracy. 
For example, at 0.01 fault rate, ResNet18 (trained on CIFAR-10 dataset) with FAQ experiences just 0.65\% accuracy drop while ResNet18 without FAQ experiences around 15.8\% accuracy drop (see label \circled{1} in Fig.~\ref{fig:Results_1}a). 
Moreover, at 0.04 fault rate, ResNet18 with FAQ experiences 3.15\% accuracy drop while ResNet18 without FAQ experiences around 79.5\% accuracy drop (see label \circled{2} in Fig.~\ref{fig:Results_1}a). 
The figure also shows that the impact of faults and the effectiveness of FAQ varies across DNNs and datasets. 
This can be observed by comparing the results marked with label~\circled{2} in Fig.~\ref{fig:Results_1}a with the results marked with label \circled{3} in Fig.~\ref{fig:Results_1}d and the results marked with label~\circled{6} in Fig.~\ref{fig:Results_1}c. However, note that in all cases, FAQ offers significant performance improvements; for example, see the difference between the accuracy of \textit{`with FAQ'} and the accuracy of \textit{`without FAQ'} at \circled{2}, \circled{3}, \circled{4}, \circled{5}, and \circled{6}. 

\begin{figure}[ht]
\centering
\includegraphics[width=1\linewidth]{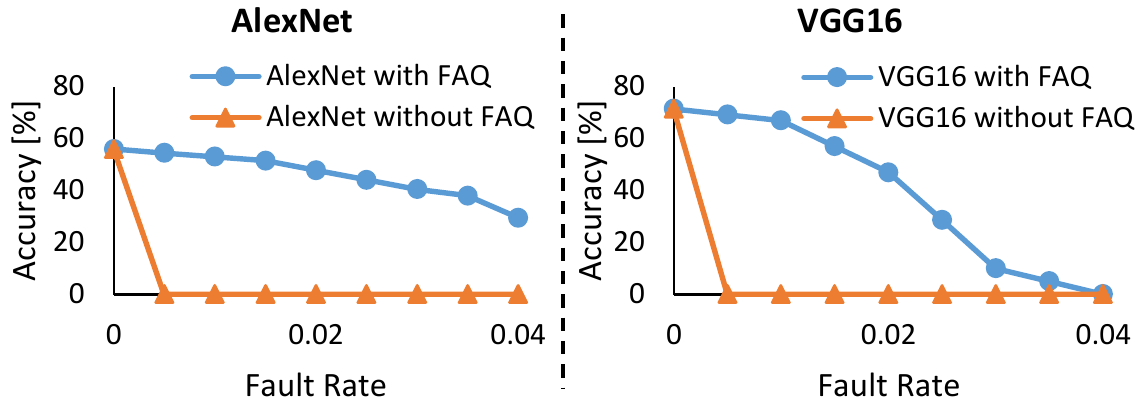}
\caption{Impact of FAQ on the accuracy of the AlexNet and the VGG16 models trained on the ImageNet dataset.}
\label{fig:Results_ImageNet}
\end{figure}

We also tested the effectiveness of FAQ for DNNs trained on the ImageNet dataset and observed similar results, i.e., FAQ offers significant improvements in DNN accuracy at low and moderate fault rates. The results are presented in Fig.~\ref{fig:Results_ImageNet} for two different DNNs. 

\subsubsection{Impact of Protected First and Last Layers}
\begin{figure}[ht]
\centering
\includegraphics[width=1\linewidth]{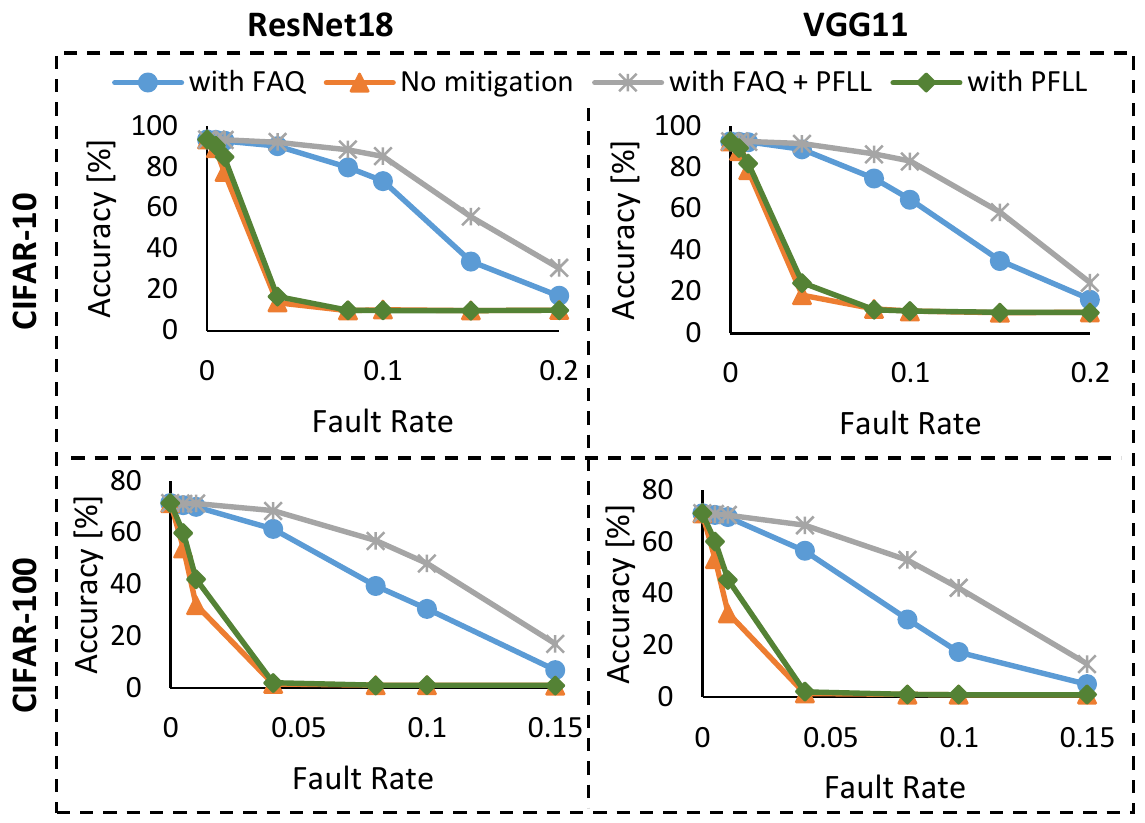}
\caption{Impact of using protection for the first and the last layer of the DNNs on their accuracy. PFLL corresponds to \textit{Protected First and Last Layer} setting. }
\label{fig:Results_PFLL}
\end{figure}

Following the trend in Quantized Neural Networks (QNNs) domain, we also simulated the case where the first and the last layer of the given DNN is executed using fault-free (protected) hardware while all rest of the intermediate layers are executed using the given faulty hardware (having stuck-at permanent faults in the on-chip weight memory). 
Fig.~\ref{fig:Results_PFLL} presents a comparison between no mitigation case, FAQ with Protected First and Last Layer (PFLL), and FAQ without PFLL. The results clearly show that for all the DNNs, adding protection for just the first and the last layer can result in a significant improvement in the DNN accuracy. 

\subsubsection{Effectiveness of FAQ for Fault-Aware Retaining} 

\begin{figure*}[ht]
\centering
\includegraphics[width=1\linewidth]{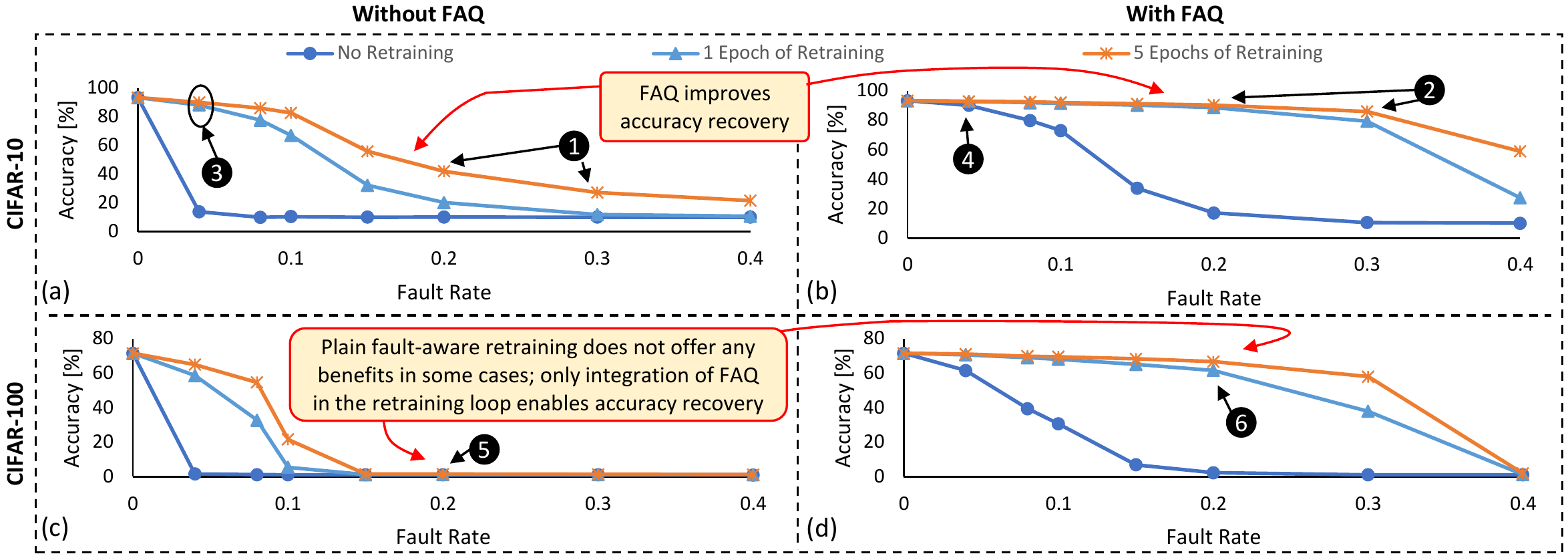}
\caption{Impact of fault-aware retraining on the accuracy of ResNet18 when used \textit{with} and \textit{without} FAQ.}
\label{fig:Results_Retraining}
\end{figure*}

Fig.~\ref{fig:Results_Retraining} highlights the impact of FAQ when used inside (together with) fault-aware retraining for the ResNet18 model. By comparing the results marked with label~\circled{1} in Fig.~\ref{fig:Results_Retraining}a with the results marked with label~\circled{2} in Fig.~\ref{fig:Results_Retraining}b, it can be concluded that using FAQ inside fault-aware retraining helps in accelerating the accuracy recovery process. 
For example, at 0.2 fault rate, after 5 epochs of fault-aware retraining integrated with FAQ, the ResNet18 model trained on the CIFAR-10 dataset (on average) reaches 90.4\% accuracy (i.e., only 2.94\% accuracy drop from the baseline) while for the same scenario but without FAQ, fault-aware retraining could push the accuracy only to 42\% (i.e., 51.35\% accuracy drop from the baseline). 

By comparing \circled{3} in Fig.~\ref{fig:Results_Retraining}a with \circled{4} in Fig.~\ref{fig:Results_Retraining}b, we observe that, in some cases, plain FAQ (i.e., without fault-aware retraining) can offer better results compared to moderate levels of plain fault-aware retraining (i.e., without FAQ). For example, in this case, at 0.04 fault rate plain FAQ offers 90.23\% accuracy (\circled{4} in Fig.~\ref{fig:Results_Retraining}b) while 1 epochs of plain fault-aware retraining offers 88.2\% accuracy and 5 epochs offers 90.13\% accuracy (\circled{3} in Fig.~\ref{fig:Results_Retraining}a). 
Apart from the above, by comparing the results marked with label \circled{5} in Fig.~\ref{fig:Results_Retraining}c with the results marked with label \circled{6} in Fig.~\ref{fig:Results_Retraining}d, we observe that, in some cases, plain fault-aware retraining maybe complete useless, and only the use of FAQ inside fault-aware retraining can lead to accuracy recovery. 

In summary, based on the above-highlighted observations and discussion, we conclude that:
\begin{enumerate}
    \item FAQ when used inside fault-aware retraining significantly boosts the accuracy recovery of DNNs subjected to faults in the on-chip weight memory of DNN accelerators.
    \item In some cases, FAQ even outperforms plain moderate level of fault-aware retraining. 
    \item In some cases, FAQ enables accuracy recovery while plain fault-aware retraining leads to no improvements. 
\end{enumerate}

\subsubsection{Execution Time Analysis}
We implemented FAQ using a lookup table-based approach. Table~\ref{tab:timing} summarizes the execution time of the lookup table generation block for 8-bit signed fixed-point representation and the time required to apply FAQ on different DNNs trained on the Cifar-100 dataset. 
For all the DNNs, the execution time of FAQ is less than 3 seconds when executed using a CPU. 
In each case, most of the time (more than 90\%) is consumed by the mask generation block. 
Compared to fault-aware retraining of ResNet18 for Cifar-100 on a machine equipped with one GTX3080, the execution time of FAQ is less than 5\% of the time required to run 1~epoch of retraining. 
\textit{In summary, it can be concluded that FAQ is a low-cost techniques which can offer significant performance gains. In cases where plain FAQ is insufficient to meet the user requirements, it can be integrated in fault-aware retraining to accelerate the accuracy recovery process.} 

\begin{table}[ht]
    \centering
    \caption{Execution time of FAQ for different DNNs trained on CIFAR-100.}
    \label{tab:timing}
    \includegraphics[width=0.7\linewidth]{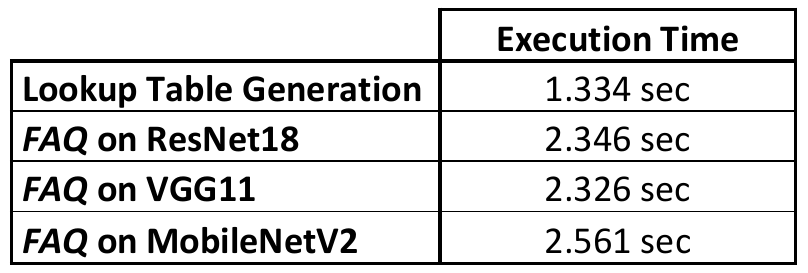}
\end{table}

\section{Usability of the Proposed \textit{FAQ} Technique}
\begin{figure}[ht]
\centering
\includegraphics[width=1\linewidth]{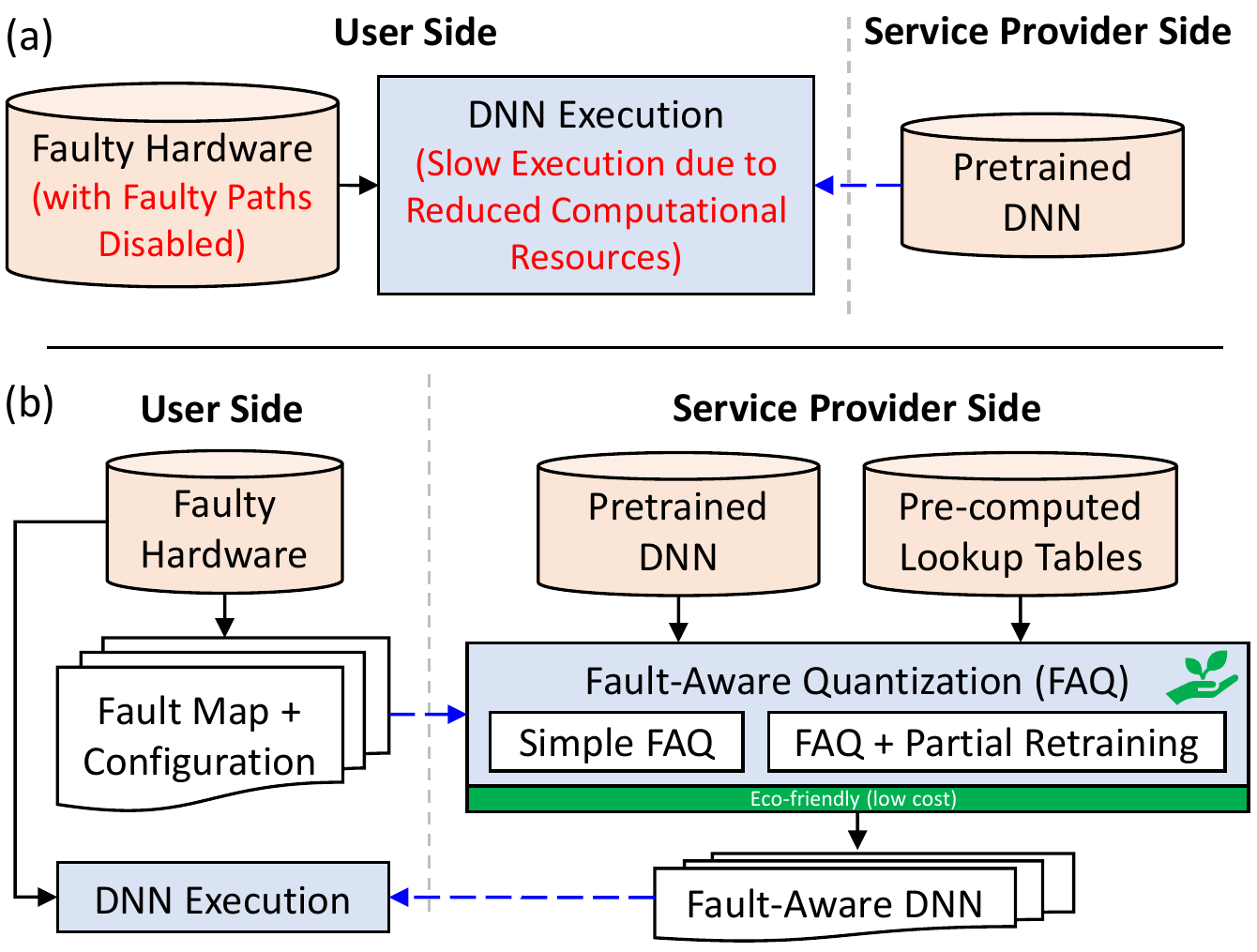}
\caption{Usability of the proposed FAQ technique: (a) \textbf{Scenario 1} highlights the \textbf{common case} where fault mitigation is achieved by disabling the faulty paths in the hardware, i.e., at the cost of lower speed or efficiency; (b) \textbf{Scenario 2} highlights the \textbf{proposed method} where FAQ-like low-cost techniques are employed to quickly tune the given DNN to offer fault mitigation without compromising speed or efficiency but at the cost of negligible accuracy loss.} 
\label{fig:Usability}
\end{figure}

Fig.~\ref{fig:Usability} presents two different scenarios related to how a faulty hardware can be employed for reliable DNN execution. 
In the first scenario (Fig.~\ref{fig:Usability}a), faulty components are disabled and the modified hardware is used to run applications \textit{but at a lower speed/efficiency}. 
This post-fabrication modification technique is commonly used by many leading industries to achieve higher yield~\cite{Some_AMD}~\cite{With_CpU}, mainly by selling the faulty chips (after post-fabrication modifications) as low-performance variants. 
On the other hand, DNNs are considered to be (to some extent) inherently resilient to errors. 
Therefore, in the second scenario (Fig.~\ref{fig:Usability}b), this error resilience of DNNs is exploited to quickly (in a resource-efficient manner) tune the required DNN for a given faulty hardware using techniques such as the proposed FAQ and offer fault mitigation without compromising speed/efficiency but at the cost of negligible accuracy loss. 
Note that DNN tuning can be performed anywhere, at the user end or at the service provider end. 
Considering the above setup, we envision that technologies like NVIDIA DLSS (Deep Learning Super Sampling) can significantly benefit from FAQ-like frameworks that offer low-cost fault mitigation to boost performance at the cost of unnoticeable quality loss. 

\section{Conclusion}
In this paper, we proposed a novel concept of Fault-Aware Quantization (FAQ) for mitigating the effects of stuck-at permanent faults in the on-chip weight memory of DNN accelerators. 
FAQ offers fault mitigation at a negligible cost compared to fault-aware retraining while offering comparable (or, in some cases, better) accuracy results. 
We proposed a lookup table-based algorithm to achieve ultra-low model conversion time. 
We demonstrated the effectiveness of the proposed FAQ technique for five different DNNs, i.e., ResNet-18, VGG11, VGG16, AlexNet and MobileNetV2, and three different datasets, i.e., CIFAR-10, CIFAR-100 and ImageNet. 
We also demonstrated the efficacy of the proposed technique for improving fault-aware retraining used for mitigating stuck-at faults in the on-chip weight memory of DNN accelerators. 
Note that the proposed technique is not limited to CMOS-based DNN accelerators. It can also be adopted for the emerging technology-based devices such as ReRAM-based DNN accelerators. 

\section*{Acknowledgment}
This work was supported in parts by the NYUAD Center for Artificial Intelligence and Robotics (CAIR), funded by Tamkeen under the NYUAD Research Institute Award CG010, the NYUAD Center for Interacting Urban Networks (CITIES), funded by Tamkeen under the NYUAD Research Institute Award CG001, and the NYUAD Center for CyberSecurity (CCS), funded by Tamkeen under the NYUAD Research Institute Award G1104.

\def\bibfont{\footnotesize} 
\bibliographystyle{IEEEtran}
\bibliography{Refs}

\end{document}